\magnification1200

\vskip 2cm
\centerline{{\bf  Memories of Abdus Salam and the early days of supersymmetry }\footnote{$\dagger$} {Invited contribution to the second edition of The Supersymmetric World, The beginnings of the Theory ed. by M. Shifman, World Scientific.}}

\vskip 1cm
\centerline{\bf   Peter West }
\vskip 1.2cm
\centerline{{ \it Mathematical Institute, University of Oxford, Woodstock Road, Oxford, OX2 6GG, UK}}
\vskip 1.2cm
\centerline{ { \it Department of Mathematics, King's College, London WC2R 2LS, UK }}
\vskip 1.7cm

\vskip 1.7cm
\centerline{\sl Abstract}
I  give an account  of what it was like to be a PhD student of Abdus Salam and also to take part during the early stages of the development of supersymmetry. 

\noindent
\vskip 4cm
email:  peter.west540@gmail.com
\vfill

\eject
\medskip
I thank Misha Shifman for the suggestion to write this article. Its purpose is not to give an account of the  life of Abdus Salam or a systematic account of the early years of supersymmetry. Rather it is to give an impression of what it was like to be a PhD student of Abdus Salam and also to take part during the early stages of the development of supersymmetry. 

After an enjoyable undergraduate degree at Imperial college I was accepted for PhD studies at Imperial  in 1973 but who would be my supervisor was not specified. I had a choice  of two possible supervisors,   Tom Kibble or Abdus Salam, or Professor Kibble and Professor Salam as us students referred to them. It was a measure  of the strength of Imperial  College in theoretical physics at that time   that a student could have a choice between two  professors  who had both already laid the cornerstones of the standard model and who had also made other very important contributions. Although at that time there was not much talk of the standard model,  at least among us students or in our lectures. Professor Kibble was very friendly and  polite but he was rather reserved and did not  say anything he had not thought carefully about. Indeed he did not say very much at all. His deep understanding of physics was obvious during the lecture course that  he gave us. 

In contrast Professor Salam was very lively,  had charisma and an enthusiasm for physics but he also had a friendly disposition which made one feel at home, at least that was the case for me. As he explained some point of physics   he usually  had a faint playful smile on his face and one could not help wanting  to understand what he was saying and join in. I got the sense that studying under Professor Salam would be very exciting and one might find results of interest. Being keen to make my way in theoretical physics I choose Professor Salam despite the fact that  I was warned that many PhD students of Professor Salam did not prosper. Ali Chamseddine also began with Professor Salam at the same time. 

For my first year I did not see much of Professor Salam as I was taking my preparatory
courses, but once the summer came I had finished my courses and so I went to see Professor
Salam to find out what I would work on. At that time Professor Salam had an office on the fifth floor of the 
physics building but later on he  moved office to the newly built Huxley Building. In the former office his 
door would be open but a large screen had been placed near to the door which blocked the view of his desk where he worked. 
While in the latter office the door would be only slightly ajar and it did not permit a view of this desk. 
A knock at the door resulted in  a soft voice  beckoning you in. 

To my surprise he asked me what I wanted to do. I said that the infinities in quantum field theory 
were very ugly and so I would like to work on general relativity which was more aesthetically pleasing. Rather than explain the
flaws in this naive approach, he suggested that there would  not be so much to do in general
relativity and that I might like to look at his very recent paper with John Strathdee in
which they had discovered superspace and also super-Feynman rules [1]. With this he handed me the ICTP preprint 
with its characteristic pale blue boarder.   As he sometimes did on later occasions  he did not say I might like to look at this humble paper. I  do not remember that he  said much about the paper but only that I should look at it. Professor  Salam had previously written papers on constructing spacetimes from cosets, for example the one with Mack [2] on viewing the conformal group from the viewpoint of a coset. As such it was natural that he and Strathdee should construct  superspace as the coset of the super Poincar\'e group modulo the Lorentz group. Within days I was captured by the ideas in this paper and began working on infinities in supersymmetric theories. From the perspective of today I realise that I had been subject to his great charm
and diplomacy, a skill which he had used to such great effect all over the world.

Professor Salam was not often at Imperial as he spent most of his time at the centre he had created in Trieste.  Little did I, as a student, know of the many endeavours for the good of science in the third world that he was undertaking. Us students never knew when we would next see Professor Salam. He would come about once a month and the first we realised that he was in was that the door of his room was slightly  ajar as we came into work. I would knock and he would welcome me in. Remarkably he was usually sitting at his desk studying some document  by himself. He was always very cheerful, friendly and seemed to have time to talk as if he had all the time in the world. Since he was away most of the time we had no one to watch over us and as a result we did not work very hard unless we really thought something was going to work out in a very good way. 

Supersymmetry had been discovered in 1971 in the very remarkable paper of Golfand and Liktman [3] 
and a non-linear realisation was  later found by Volkov and Akulov [4]. These  papers went essentially unnoticed even though the former paper   did contain a four dimensional model of supersymmetry.  Supersymmetry also appeared in the superstring whose 
fermionic sector was  found by Ramond [5] and  the bosonic sector in the later paper of 
Neveu and Schwarz [6]. Hidden in these papers was the two dimensional supersymmetry algebra and this was extracted by Gervais and Sakita [7].  Wess and Zumino generalised this algebra to four dimensions and constructed the invariant Wess-Zumino model [8]. 

It was only with this last paper that a few physicists took an interest in supersymmetry. They were located at a only a few
institutions including CERN in Geneva, the Ecole Normale Sup\'erieur in Paris, Karlsruhe  and the two institutes to which Salam belonged, namely Imperial College and the International Centre for Theoretical Physics in Trieste. 
Thus in 1973 there were only a  few  people working on supersymmetry. Being a  student of Salam I was one of them as were indeed a number of other young people who were the PhD students of those few who thought that supersymmetry was of interest. 

In trying to read the  paper of Salam and Strathdee it quickly became  clear that supersymmetry required 
very advanced manipulations in the matrix algebra associated with fermions. There were no courses on supersymmetry in 1973, at least not at Imperial College,  nor were there any papers where one could  look  this,  and a number of other difficult matters,  up. 
 Fortunately Professor Salam's long term collaborator Bob Delbourgo had an office nearby and he provided me with all the technical help I needed. He even had a problem I might  work on involving  the calculation of  superspace Feynman diagrams in the Wess-Zumino model. Understandably he was unable to resist the temptation do this himself  and in a matter of days he had found the result [9] before I had even begun to think about the problem.  Using the superspace Feynman rules he found  that the Wess-Zumino model only needed a wavefunction renormalisation as had previously been found using the component formalism by   Wess and Zumino [8] and Iliopoulos and Zumino [10]. 

At that time the big problem was to spontaneously break supersymmetry  and so hopefully find a realistic
model of nature that was supersymmetric. Particles in a supersymmetric theory had the
same mass and although one could spontaneously break supersymmetry at the classical
level [11,12], the pattern of masses this lead to was not consistent with nature. Professor
Salam thought that radiative corrections would break supersymmetry and lead to more
promising results. Such corrections had been used to good effect in a paper of Coleman and Weinberg [13]  in  the  very different context, namely scalar electrodynamics. 

Professor Salam and  his long term collaborator John Strathdee, who was in Trieste, produced
a series of models and it was my job to compute their one-loop effective potentials and
see if supersymmetry was spontaneously broken and what pattern of masses they lead to.
The first models did not work and as time went on the models became more and more
complicated involving very many fields. If I had not completely finished computing with
a given model by the time I next meet with Professor Salam it was not a problem, there
was always a much better model to look at instead. 

Eventually Bob Delbourgo and I  looked at one of the proposed  models that required us to diagonalise a 26 by 26 matrix. After much work we did do this but the entires were not real numbers meaning that the minimum of the potential which we had  also given was not actually a minimum. When I told Professor Salam about this result he was not at all perturbed and quickly produced another model which he must have brought with him from Trieste. Fearing the same fate for this model I suggested there was a problem with the model to which Salam gave a quick reply. It was not clear who was giving the worst argument. 
Shortly afterwards Salam spoke to Bob Delbourgo and ask him to tell me that all was well.

Eventually I realised that if supersymmetry was preserved at the classical level then
the effective potential vanished in the most general ${\cal N} = 1$ theory invariant under rigid
supersymmetry [14]. This meant that one could not spontaneously break supersymmetry using 
perturbative quantum corrections, although one could still hope that it was
spontaneously broken by non-perturbative corrections. The problem of breaking supersymmetry 
in a natural way is still largely unsolved. Unlike in most theories the  minimum of potentials in supersymmetric theories do not generically fix all the values of the fields and so there is a degeneracy.  Since the quantum corrections to the  potential vanish  
if supersymmetry was preserved this degeneracy was not determined by quantum corrections. 
Thus supersymmetric theories have moduli.  
\par
The result had another more favourable
consequence, as was pointed out by others [15], namely that supersymmetry did solve the
hierarchy problem, at least technically. In a supersymmetric version of the standard model
the Higgs mass would not be swept up to some large unified scale by quantum corrections
as long as supersymmetry was not broken much above the weak scale. This in turn lead
to the hope that supersymmetry might be found at the LHC (Large Hadron Collider at CERN). My article in this volume gives a 
more complete account of these developments.  As we now know the LHC did not find supersymmetry, at least not so far. While the above argument was plausible there is nothing wrong with supersymmetry being broken at a very high scale and the problem of naturalness in the standard model is resolved in some other, as yet unknown, way. A more general non-renormalisation theorem was later found by Grisaru, W. Siegel and  M. Rocek [16] using some reformulated superFeynman rules. 

In those days there was no internet and so no arXiv. Once a result was written up the paper was sent  to the journal but it was usually a long time before it would be published even if the referees were favourable. To enable more speedy access to  the paper  copies of it,  called  preprints, were produced  and these were  sent out by mail to those institutes whose members might be interested in it. Since there were not many people who were working on supersymmetry this did not involve much effort. People who had heard of the result and who did not have access to  the preprint would send a post card requesting a copy which was then sent directly to them. Each week the 
preprints that had arrived at Imperial would be placed on some racks in the corridor. Each paper had a piece of paper attached  to it on which one could write ones name which resulted in it being passed to you but only after the others who were higher up the list had read it. The consequence was that people working on supersymmetry were not always aware of what every one else was doing. One advantage was that people could pursue  lines of independent research without too much influence of what others were doing. 

After three years I wrote up my results in a thesis and the time came for my PhD examination. 
Professor Salam had enrolled Professor John Taylor from  Cambridge University  to be the external examiner. He was rather softly spoken, very polite  and had found important results in how symmetries were realised in quantum field theory. 
They  assembled in Salam's office and I was called in. Professor Salam indicated that John Taylor should ask me some questions on  my thesis which he then did, whereupon Professor Salam began working at his desk on some 
no doubt important task. This lasted half an hour at most and then there was silence which Professor Salam 
eventually noticed. He turned to me and said I should explain to Professor Taylor some results on gauging the super 
Poincar\'e group  I had found with Ali Chamseddine. With this Professor Salam resumed the study of the document he was reading and  I began writing  on  the board. This  lasted at most  twenty minutes whereupon silence returned which Professor Salam again noticed. At this point the exam was declared over and, after a glance between them,  they told me I had passed. 

After the Wess-Zumino paper an obvious problem was to find a theory of gravity which was supersymmetric, that is, a generalisation of Einstein's theory to contain spin two as well as a spin three-halves particle. There was an attempt [17] to construct supergravity by generalising  the geometry of Einstein's theory  but  it was not clear how to do this. 
After about two years the first supergravity theory  was found by Ferrara,  Freedman and van Nieuwenhuizen [18]. They used the low key,  but very powerful Noether technique that had been discussed in the context of gravity by others. The advantage of this approach was that it did not require any preconceived geometrical notions,  the disadvantage was that it required a lot of difficult algebra. Indeed they were only able to show that the supergravity theory they had found was invariant under supersymmetry by using a computer. The problem was that the  result could not be reproduced by others. Deser and Zumino then  showed that the spin three-halves particle did, unlike in other theories, propagate in a causal manner and they  also discussed the theory from the viewpoint of first order formalism [19]. 

There then followed two papers coupling the new supergravity theory to the matter supermultiplet containing  spin one and spin one half  particles [20]. The next  paper on supergravity was the result of Ali Chamseddine and myself [21] which Salam had asked me to explain to Professor Taylor.  We  had taken  it into our heads to gauge the super Poincar\'e group in the style of Yang and Mills. Of course even Einstein's gravity is not a Yang-Mills theory  but we were students who did not know  any better. After only  a few lines many of the features of the  new supergravity theory emerged. Indeed one could write down an action which was just that found in the paper of Ferrara,  Freedman and van Nieuwenhuizen. One great advantage was that one could prove using the usual algebraic manipulations,   and in a relatively few lines,  that the supergravity action was invariant under the supersymmetry transformations. In this way the discovery of supergravity theory was completed and  it was in this way that the  all subsequent supergravity theories were shown to be invariant. Our approach had a significant draw back,  it required one to adopt a condition that broke the gauge symmetry, but miraculously once one took this  unconventional step everything worked very well and the desired result emerged. Professor Kibble suggested in his understated way that we should understand this step better. Our proof of the invariance of the action was further elucidated in a paper by Paul Townsend and van Niewenhuizen [22]. 

Of course we had also derived Einstein gravity as a gauge theory of the Poincar\'e group. The gauging technique was used to find the conformal supergravity theories  [23] and higher spin theories emerge if one took a much bigger gauge group. Ali Chamseddine and myself also gauged the OSp(1,4)  group and found the transformations of the vierbein, spin connection and gravitino. The results were reported in thesis of Ali Chamseddine [24]. An action was later found by  MacDowel and Mansouri  [25] which agreed with the then known supergravity theory with a cosmological constant [26].  One problem was that our paper was rejected by Nuclear Physics B and being students we did not know what to do next. As a result  it took quite some time to resubmit it after removing some of the fibre bundle language which really had no proper place in the paper. An account of previous  work on gauge symmetry and gravity by Sciama, Kibble,  Trautman and others and the connection to our work  is explained in detail in reference [27]. 

The time came to get a job. I applied to the research council of the UK for  a post doctoral fellowship. The resulting interview did not go well. The panel did not think much of the new fangled idea of supersymmetry and wanted to know how it would affect current particle physics experiments. I, nor anyone  else, had any real idea of what was the answer to this question and I did not get a fellowship. The Royal Society did offer me a fellowship to take to the Ecole Normale Superieur. No doubt I had Salam and others at Imperial to thank for this. Most  of the young people who worked on supersymmetry had to wait a long time to gain a permanent job although I had better luck. 

There was another problem with the original supergravity theory found by Ferrara,  Freedman and van Nieuwenhuizen. 
It had a supersymmetry algebra which was tied to the supergravity theory rather than being an independent algebra. 
Until this point a symmetry of a theory  existed in a way that was dependent of that theory.  In other words the transformations formed a group which had an independent life.  For example,  the Poincar\'e transformations exist independent  of Maxwell's theory and so can be used as a guide to construct other Poincar\'e  invariant theories. 
As such to construct the coupling of super matter to the original supergravity was very difficult since this required a new supersymmetry algebra and this was only achieved for a few  very specific models [20,28]. 

What to do was not clear for about two years. The key step was the observation that the fields of a supergravity theory were in a one to one  correspondence with the currents in a supersymmetric theory and this had already been worked out by Ferrara and Zumino [29]. This supermultiplet  contained the energy momentum tensor and the supercurrent but there were other objects which had to correspond to additional fields in the supergravity theory. To find the theory we used the same Noether procedure. It was a very long but enjoyable calculation. The  extension of the original supergravity theory which did have a supersymmetry algebra that was independent of the supergravity theory was found by Kelly Stelle and myself [30] and independently by Ferrara and van Nieuwenhuizen [31]. This theory had additional fields that did not lead to physical degrees of freedom. It was then possible  to construct a tensor calculus for supergravity [32] [33] and then it was relatively easy to find the most general coupling of matter to supergravity  [34,35,36] which played such a key role in the construction of realistic models of supersymmetry using the ideas of Chamseddine, Arnowitt  and Nath   [34, 37]. Other ways to find the most general coupling of matter to supergravity  have proliferated but they perhaps lack the simplicity  and efficiency of the original approach. 

Perhaps on the basis of these results Professor Salam invited me to visit him in Trieste for sometime.  When I arrive at the centre in the morning  I would find Professor Salam sitting at his desk studying what I thought  were the latest papers.  Talking to Professor Salam you could not escape his great enthusiasm for physics; you came to understand that it was a lot of fun to do physics and that it was good to work in a very relaxed and free thinking way. Once we began our discussion he would think of a vast number of ways to proceed in the quest to find new things. He was always most interested in very new ideas and while not all of his ideas worked they included many of the deepest ideas that have come to dominate the subject. As one of his students I was, perhaps, able to absorb some of these qualities. Certainly, it was due to him that I began working on supersymmetry
rather than on some uninteresting direction.

At the end of my visit to Trieste the time came for me to leave for the airport. Salam
realised that I would be travelling at the same time as the Italian Minister for Science, who
was visiting the centre, and so he suggested we could share the same car to the airport.
This was met with a frown by the organiser of the visit who, no doubt correctly, thought
that a scruffy post-doc with a rucksack might dent the carefully created image that the
centre wanted to portray. Salam's suggestion was quietly shelved and  I went by train to the airport.

I did not meet Julius Wess or Bruno Zumino during my doctorate studies but their presence 
was always with me and, I suspect,  the other young physicists working on supersymmetry. Their results and the  clarity of their papers was a benchmark to which we could but try to attain.  When we did meet  they were very friendly and encouraging.  One aspect of there being so few working on supersymmetry was that there was a feeling of a common purpose and being enguaged in an exciting endeavour. Of course supersymmetry was great to work on. The calculations were usually  difficult but often worked out  in a spectacular manner and this encouraged us to continue despite the lack of support from the community at large. 
 
 The superspace formulation of supergravity using geometrical concepts analogous to those used in Einstein's theory was found by Wess and Zumino in [38]. It turned out that the curvatures and torsion were subject to very strong constraints which were allowed by the fact that the tangent space group was only the Lorentz group under which they  formed a very  reducible representation. We suspected that they had used the fact that it must lead, when expressed in components,   to above formulation of supergravity to guide them to this result. There then appeared a paper by  Warren Siegel claiming that their constraints were so strong that there were no physical degrees of freedom left [39]. Wess and Zumino explained that this paper was wrong but  they had a worried look. Of course we found it  very amusing that a young unknown, at least  to us,  post doc had dared to contradict a  paper of the gods.  Eventually it was clear that the paper of Warren Siegel was indeed wrong but he then wrote a paper [40] solving the constraints on  the  torsions and curvatures. A pedagogical and further elucidation of  this work was given in a paper by Gates and Siegel [41]. It was a very considerable feat involving technical leaps of such difficulty that I suspect few have worked through it. 
 
 The last time I meet Bruno Zumino was  in the Newton Centre  at  the University of Cambridge in 1997.  David Olive and I were  puzzled by a point to do with a coset construction and we discussed it with Bruno late in the afternoon. It was not a crucial point but it would have been good to understand it. The next day when I arrive quite early  I found a hand written note on my desk from Bruno which had worked it all out in great detail and in a very clearly way. David also found a note on his desk.

 I end with  an account of two meetings with Abdus Salam that display his warmth
and humanity.

I met Abdus Salam in his office in London a few days after he had won the Nobel
prize. I asked him what was it like to win such a prize, he reassured me that he was just
the same. He then suggested that we go for coffee in the common room in the old physics
building at Imperial. To get there we had to go through a number of doors and he insisted
that I go first through each door despite my protests.

During the time that Salam was very ill there was a conference in his honour at Trieste,
but he was not well enough to go to all the talks. I saw him sitting at the very back of the
big auditorium. I asked if it would be alright to say hello, but I was told that he might
not recognise me. Since this might be the last time I would see him I went anyway. I said
hello, he put up his hand and I shook it. He then immediately said how was Sue. Sue is
my wife's name who he had meet only once many years before.

\medskip
{\bf Acknowledgements}
\medskip 
I want to thank Misha Shifman for suggesting I write this article. I also thank Ali Chamseddine for comments. 
\medskip
{\bf References}
\medskip
\item{[1]} A. Salam and J. Strathdee, {\it On Superfields and Fermi-Bose Symmetry}, Phys. Rev.
{\bf D11} (1975) 1521.
\item{[2]} G. Mack and A. Salam, {\it Finite-Component Field Representations of the Conformal Group}, Annals of Physics {\bf 53}, (1969), 174-202. 
\item{[3]} Y.A. Golfand and E.S. Likhtman, {\it Extension of the Algebra of Poincar\'e Group Generators and Violation of P Invariance}, {\it JETP Lett.} {\bf13}, 323 (1971). 
\item{[4]} D.V. Volkov and V.P. Akulov, {\it Possible universal neutrino interaction},  {\it Pis'ma Zh. Eksp.
Teor. Fiz.} {\bf16}, 621 (1972); {\it Phys. Lett.} {\bf46B}, 109 (1973). 
\item {[5]} ÊÊP. Ramond, {\it Dual theory for free fermions}, Phys. Rev. {\bf ÊD3} (1971) 2415.
\item {[6]} A. Neveu and J.H. Schwarz, {\it Factorizable dual model of pions}, Nucl. Phys. {\bf B31} (1971) 86;  A. Neveu and J.H. Schwarz, {\it Quark Model of Dual Pions}, Phys. Rev.{\bf  D4} (1971) 1109. 
\item {[7]} J. L. Gervais and B. Sakita, {\it Field Theory Interpretation Of Supergauges In Dual Models}, Nucl. Phys. {\bf B34} (1971)
632.
\item{[8]} J. Wess and B. Zumino, {\it Supergauge transformations in four dimensions}, Nucl. Phys.
{\bf B70} (1974) 139; A Lagrangian Model Invariant Under Supergauge Transformations,
Phys. Lett. {\bf 49B} (1974) 52.
\item{[9]} R. Delbourgo, {\it Superfield Perturbation Theory and Renormalization},     Nuovo Cim. {\bf A 25} (1975) 646. 
\item{[10]} J. Iliopoulos and B. Zumino, {\it Broken supergauge symmetry and renormalization}, Nucl. Phys. B76 (1974) 310. 
\item{[11]} L. O'Raifeartaigh, {\it Spontaneous Symmetry Breaking for Chiral Scalar Superfields}, Nucl.
Phys. {\bf B96} (1975) 331.
 \item{[12]} P. Fayet, {\it Spontaneous Supersymmetry Breaking Without Gauge Invariance},  Phys. Lett.
{\bf 58B} (1975) 67.
\item{[13]} S. Coleman and E. Weinberg. {\it Radiative Corrections as the Origin of Spontaneous Symmetry Breaking},  Physical Review {\bf D7} (1973) 1888. 
 \item{[14]} P. West, {\it Supersymmetric Effective Potential}, Nucl. Phys. {\bf B106} (1976) 219.
\item{[15]} E. Witten, {\it Dynamical Breaking of Supersymmetry}, Nucl. Phys. {\bf B188} (1981) 513.
\item{[16]}  M  Grisaru, W. Siegel and  M. Rocek,   {\it Improved Methods for Supergraphs}, Nucl.Phys.B {\bf  159} (1979) 429, 
\item{[17]} P. Nath and R. Arnowitt, Phys lett {\bf 56B }(1975) 177, R. Arnowitt, P. Nath and B. Zumino, Phys. Lett {\bf 56B}, (1975) 81, 
 \item{[18]} D. Freedman, P. van Nieuwenhuizen and S. Ferrara,  {\it Progress toward a theory of supergravity }, 
Phys. Rev. {\bf D13}, 3214 (1976); D. Freedman and P. van Nieuwenhuizen,  {\it Properties of supergravity theory}, Phys. Rev. {\bf D14}, 912 (1976). 
\item {[19]} S. Deser and B. Zumino,  {\it Consistent Supergravity}, {\it Phys. Lett.} {\bf 62B}, 335 (1976). 
\item{[20]} S. Ferrara, J. Scherk and P. van Nieuwenhuizen, {\it Locally Supersymmetric Maxwell-Einstein Theory},  {\it
Phys. Rev. Lett.} {\bf 37}, 1035 (1976); S. Ferrara, F. Gliozzi, J. Scherk and P. van Nieuwenhuizen, {\it Matter Couplings in. Supergravity Theory}, {\it Phys. Rev.} {\bf D15}, 1013
(1977). 
\item{[21]} A. Chamseddine and P. West, {\it Supergravity as a gauge theory of supersymmetry}, {\it Nucl. Phys.} {\bf
B129}, 39 (1977). This paper was received on the 28 September 1976 but was initially rejected for publication. The  revised version,  received  on  10 June 1977, was essentially the same except for two paragraphs explaining in more detail its relation to references [18] and [19]. It was also published as the  Imperial preprint ICTP/75/22, September 1976. 
\item {[22]} P. Townsend and P. van Nieuwenhuizen, {\it Geometrical interpretation of extended supergravity},  {\it Phys.
Lett.} {\bf B67}, 439 (1977). 
\item{[23]} P. Van Nieuwenhuizen, M. Kaku, and P. Townsend, {\it Properties of conformal supergravity}, Phys. Rev. {\bf D17}, (1978) 1501. 
\item{[24]} A. H. Chamseddine, {\it Supersymmetry and higher spin fields},  Ph.D. Thesis, defended September 1976, Imperial College, London University. Chapter 6 of the thesis is stated as collaborative work with P. West. 
The thesis can be found at 
\par
https://drive.google.com/file/d/0B8tITtoqQkxfeHQ3cFJ0WUxIcHM/view?usp=sharing \ 
\item{[25]} S. MacDowell and F. Mansouri, {\it Unified geometric theory of gravity and supergravity}, Phys. Rev. Lett. {\bf 38 }(1977) 739. This paper was received on the 9 February 1977. 
\item{[26]} Paul K. Townsend, {\it Cosmological constant in supergravity}, Phys. Rev. {\bf D15} (1977) 2802. 
\item{[27]} A. Chamseddine and P. West, {\it The role of the 1.5 order formalism and the gauging of  spacetime groups   in  the development of   gravity and supergravity theories}, Mod.Phys.Lett. {\bf A 37} (2022) 08, 2230005, arXiv:2201.06874. 
\item{[28]}  P. Breitenlohner, D.Z. Freedman, {\it Supergravity with axial-gauge invariance},  {\it Phys. Rev.} {\bf D15}, 1173 (1977). 
\item{[29]} S. Ferrara and B. Zumino, {\it  Transformation properties of the supercurrent}, Nuclear Physics {\bf B87}(1975) 207
\item{[30]} K. Stelle and P. West, {\it Minimal auxiliary fields for supergravity}, {\it Phys. Lett.} {\bf B74}, 569. 
330 (1978).   
\item{[31]} S. Ferrara and P. van Nieuwenhuizen, {\it The auxiliary fields of supergravity}, Phys.
Lett.{\bf B74}, 333 (1978).
\item{[32]}  K.S. Stelle and P. West, {\it Tensor calculus for the vector multiplet coupled to supergravity},  {\it Phys. Lett.} {\bf 77B},
376 (1978);  {\it Relation between vector and scalar multiplets and gauge invariance in supergravity},  Nucl. Phys. {\bf B145} (1978) 175.
\item{[33]} S. Ferrara and P. van Nieuwenhuizen, {\it Tensor calculus for supergravity}, Phys. Lett. {\bf 76B} (1978) 404;  
\item{[34]} {\it Structure of supergravity},  Phys. Lett. {\bf 78B}, 573 (1978). 
A. H. Chamseddine, R. Arnowitt and P. Nath, {\it Locally 
supersymmetric grand unification}, Phys. Rev. Lett. {\bf 49 }(1982) 970.
\item{[35]}  E. Cremmer, S. Ferrara, L. Girardello and A. Van Proeyen,
{\it Coupling supersymmetric Yang-Mills gauge theories to supergravity}, Phys. Lett. {\bf 116B} (1982) 231.
\item{[36]}  P. Nath, R. Arnowitt and A. H. Chamseddine, {\it Applied N=1
Supergravity, }ICTP\ series in Theoretical Physics, Volume 1, (1982), World
Scientific, Singapore.  
\item{[37]}  P. Nath, R. Arnowitt and A. H. Chamseddine, {\it Gauge
hieirarchy in supergravity in supergravity Guts}, Nucl. Phys. {\bf B227} (1983) 121.
\item{[38]}  J. Wess and B. Zumino, {\it Superspace formulation of supergravity}, Phys Lett {\bf 66B} (1971) 361
\item{[39 ]} W. Siegel,  {it A Comment on the Wess-Zumino Formulation of Supergravity} 1977, Harvard University preprint.  \item{[40]} W. Siegel, Solution to Constraints in {Wess-Zumino} Supergravity Formalism,     Nucl.Phys. {\bf B142} (1978) 301-305; 
 {\it Supergravity Superfields Without a Supermetric}, Harvard preprint HUTP-771 A068, {\it Nucl. Phys.} {\bf
B142}, 301 (1978); 
\item{[41]} S.James Gates, Jr and W. Siegel, {\it Understanding Constraints in Superspace Formulations of Supergravity},    Nucl.Phys. {\bf B163} (1980) 519.

\end